# Fundamental Limit of Nanophotonic Light-trapping in Solar Cells

Zongfu Yu, Aaswath Raman, and Shanhui Fan*

Ginzton Laboratory, Stanford University, Stanford, 94305

**Establishing the fundamental limit of nanophotonic light-trapping schemes is of paramount importance and is becoming increasingly urgent for current solar cell research. The standard theory of light trapping demonstrated that absorption enhancement in a medium cannot exceed a factor of $4n^2/\sin^2\theta$, where $n$ is the refractive index of the active layer, and $\theta$ is the angle of the emission cone in the medium surrounding the cell. This theory, however, is not applicable in the nanophotonic regime. Here we develop a statistical temporal coupled-mode theory of light trapping based on a rigorous electromagnetic approach. Our theory reveals that the standard limit can be substantially surpassed when optical modes in the active layer are confined to deep-subwavelength scale, opening new avenues for highly efficient next-generation solar cells.**

*Introduction.* The ultimate success of photovoltaic (PV) cell technology requires great advancements in both cost reduction and efficiency improvement. An approach that simultaneously achieves these two objectives is to use light-trapping schemes. Light trapping allows cells to absorb sunlight using an active material layer that is much thinner than the material's intrinsic absorption length. This then reduces the amount of materials used in PV cells, which cuts cell cost in general, and moreover facilitates mass production of PV cells that are based on less abundant materials. In addition, light trapping can improve cell efficiency, since thinner cells provide better collection of photo-generated charge carriers [1].

The theory of light trapping was initially developed for conventional cells where the light-absorbing film is typically many wavelengths thick [2-3]. From a ray-optics perspective, conventional light trapping exploits the effect of total internal reflection between the semiconductor material (such as silicon, with a refractive index $n=3.5$) and the surrounding medium (usually assumed to be air). By roughening the semiconductor-air interface (Fig. 1a), one randomizes the light propagation directions inside the material. The effect of total internal reflection then results in a much longer propagation distance inside the material and hence a substantial absorption enhancement. For such light-trapping schemes, the standard theory, first developed by Yablonovitch [2], shows that the absorption enhancement factor has an upper limit of $4n^2/\sin^2\theta$ [2-3], where $\theta$ is the angle of the emission cone in the medium surrounding the cell. This limit will be referred to in this paper as the Yablonovitch limit or the $4n^2$ limit, since we will be primarily concerned with structures with $\theta \approx \pi/2$, which has a near-isotropic emission cone.

For nanophotonic films with thicknesses comparable or even smaller than wavelength scale, the ray-optics picture and some of the basic assumptions in the conventional theory are no longer applicable. As a result, whether the Yablonovitch limit still holds in the nanophotonic regime remains an open question that is of fundamental importance. In this context, we note that none of the recent experimental [4-8] and numerical works [9-16] observed substantial improvement beyond the Yablonovitch limit over a broad solar wavelength range.

In this article, we develop a statistical coupled-mode theory that describes light trapping in general from a rigorous electromagnetic perspective. Applying this theory, we show that the $4n^2$ limit is only correct in bulk structures. In the nanophotonic regime, the absorption enhancement factor can go far beyond $4n^2$ with proper design. As a specific example, we numerically demonstrate a light-trapping scheme with an absorption enhancement factor of $12\times 4n^2$ over a virtually unlimited spectral bandwidth and with near-isotropic angular response.

*Theory.* As a concrete example to illustrate our theory, we consider a high-index thin-film active layer with a high reflectivity mirror at the bottom, and air on top. Such a film supports guided optical modes. In the limit where the absorption of the active layer is weak, these guided modes typically have a propagation distance along the film that is much longer than the thickness of the film. Light trapping is accomplished through coupling incident plane-waves into these guided modes, by introducing either a grating with periodicity *L* (Fig. 1b) or random Lambertian roughness (Fig. 1a). As a standard procedure for the theoretical study of random systems in general[17], the behavior of a system with random roughness can be understood by taking an appropriate $L \to \infty$ limit of the periodic system. Thus, without loss of generality we will focus on the periodic system in the development of our theory. As long as *L* is chosen to be sufficiently large, i.e. at least comparable to the free-space wavelength of the incident light, for each incident plane-wave, one can find a guided mode such that the grating phase-matches these modes. In the presence of such a grating the guided mode can couple to the incident wave, creating a guided resonance [18].

A typical absorption spectrum for such a film [9-11,15] is reproduced in Fig. 1c. The absorption spectrum consists of a collection of sharp peaks, each corresponding to a particular guided resonance. The absorption is strongly enhanced in the vicinity of each resonance. However, compared to the broad solar spectrum, each individual resonance has very narrow spectral width. Consequently, in order to enhance the absorption over a substantial portion of the solar spectrum, one will have to rely upon a collection of these peaks. Motivated by this observation, we develop a statistical temporal coupled mode theory that describes the aggregate contributions from all resonances.

We start by identifying the contribution of a single resonance to the total absorption over a broad spectrum. The behavior of an individual guided resonance, when excited by an incident plane wave, is described by the temporal coupled mode theory equation [19-20]:

$$\frac{d}{dt}a = (j\omega_0 - \frac{N\gamma_e + \gamma_i}{2})a + j\sqrt{\gamma_e}S \qquad (1)$$

Here $a$ is the resonance amplitude, normalized such that $|a|^2$ is the energy per unit area in the film, $\omega_0$ is the resonance frequency, and $\gamma_i$ is the intrinsic loss rate of the resonance due to material absorption. $S$ is the amplitude of the incident plane wave, with $|S|^2$ corresponding to its intensity. We refer to a plane wave that couples to the resonance as a *channel*. $\gamma_e$ is the leakage rate of the resonance to the channel that carries the incident wave. In general, the grating may phase-match the resonance to other plane-wave channels as well. We assume a total of $N$ such channels. Equivalent to the assumption of isotropic emission as made in Ref. [2], we further assume that the resonance leaks to each of the $N$ channels with the same rate $\gamma_e$. Under these assumptions, the absorption spectrum of the resonance is [19]

$$A(\omega) = \frac{\gamma_i \gamma_e}{(\omega - \omega_0)^2 + (\gamma_i + N\gamma_e)^2/4} \qquad (2)$$

For light trapping purposes, the incident light spectrum is typically much wider than the linewidth of the resonance. When this is the case, we characterize the contribution of a single resonance to the total absorption by a *spectral cross-section*:

$$\sigma = \int_{-\infty}^{\infty} A(\omega) d\omega \qquad (3)$$

Notice that spectral cross-section has units of frequency, and has the following physical interpretation: For incident spectrum with a bandwidth $\Delta\omega \gg \sigma$, a resonance contributes an additional $\sigma/\Delta\omega$ to the spectrally averaged absorption coefficient.

For a single resonance, from Eqs. (2) and (3), its spectral cross-section is:

$$\sigma = 2\pi\gamma_i \frac{1}{N + \gamma_i/\gamma_e} \qquad (4)$$

which has a maximum value of

$$\sigma_{max} = \frac{2\pi\gamma_i}{N} \qquad (5)$$

in the *over-coupling* regime when $\gamma_e \gg \gamma_i$. We emphasize that the requirement to operate in the strongly over-coupling regime arises from the need to accomplish broad-band absorption enhancement. In the opposite narrow-band limit, when the incident radiation is far narrower than the resonance bandwidth, one would instead prefer to operate in the critical coupling condition by choosing $\gamma_i = N\gamma_e$, which results in $(100/N)\%$ absorption at the resonant frequency of $\omega_0$. The use of critical coupling, however, has a lower spectral cross-section and is not optimal for the purpose of broad-band enhancement. We also note that the intrinsic decay rate $\gamma_i$ differentiates between the two cases of broad-band and narrow-band. For light trapping in solar cells, we are almost always in the broad-band case where the incident radiation has a bandwidth $\Delta\omega \gg \gamma_i$.

We can now calculate the upper limit for the absorption by a given medium, by summing over the maximal spectral cross-section of all resonances:

$$A_T = \frac{\sum \sigma_{max}}{\Delta \omega} = \frac{2\pi \gamma_i}{\Delta \omega} \frac{M}{N} \tag{6}$$

where $M$ is number of resonances within the frequency range $\Delta \omega$. In the over-coupling regime, the peak absorption from each resonance is in fact relatively small; therefore the total cross section can be obtained by summing over the contributions from individual resonances. In addition, we assume the medium is weakly absorptive such that single-pass light absorption is negligible.

Eq. (6) is the main result of this paper. In the following discussion, we will first use Eq. (6) to reproduce the well-known Yablonovitch limit, and then consider a few important scenarios where the effect of strong light confinement becomes important.

*Light-trapping in bulk structures.* We first consider a structure with period $L$ and thickness $d$ that are both much larger than the wavelength. In this case, the resonance can be approximated as propagating plane waves inside the bulk structure. Thus, the intrinsic decay rate for each resonance is related to material's absorption coefficient $\alpha_0$ by $\gamma_i = \alpha_0 \frac{c}{n}$.

The number of resonances in the frequency range $[\omega, \omega + \delta \omega]$ is

$$M = \frac{8\pi n^3 \omega^2}{c^3} \left(\frac{L}{2\pi}\right)^2 \left(\frac{d}{2\pi}\right) \delta \omega . \tag{7}$$

Each resonance in such frequency range can couple to channels that are equally spaced by $\frac{2\pi}{L}$ in the parallel wavevector $k_{//}$ space (Fig. 2a). Moreover, since each channel is a propagating plane wave in air, its parallel wavevector needs to satisfy $|k_{//}| \leq \omega/c$. Thus, the number of channels is:

$$N = \frac{2\pi\omega^2}{c^2}\left(\frac{L}{2\pi}\right)^2. \tag{8}$$

From Eq. (6), the upper limit for the absorption coefficient for this system is then $A_T = 4n^2\alpha_0 d$. To obtain the upper limit for the absorption enhancement factor $F$, we compare the absorption to the single-pass absorption, to obtain:

$$F \equiv \frac{A_T}{\alpha_0 d} = 4n^2 \tag{9}$$

which reproduces the Yablonovitch limit.

The derivation above illustrates that systems of random roughness can in fact be understood by taking the $L \to \infty$ limit. Moreover, while the derivation here is appropriate for an isotropic emission situation, the theory can be generalized to the case of an anisotropic emission cone, and reproduces the standard result (see supplementary information).

The analysis here also indicates that the Yablonovitch limit is only applicable under the condition where both the periodicity and the thickness are much larger than the wavelength. Eq. (8) is not applicable in the case where the periodicity is comparable to the wavelength, while Eq. (7) is not valid in the case where the film thickness is much smaller than the wavelength. In both cases, as we now show, substantial enhancement over the Yablonovitch limit can occur.

*Light-trapping in structures with wavelength-scale periodicity.* When the periodicity $L$ is comparable to the wavelength $\lambda$, the discrete nature of the channels becomes important (Fig. 2a). To illustrate this effect, we assume that the film has a high refractive index (for example, silicon), such that the wavelength in the material is small compared with the periodicity, and has a thickness of a few wavelengths. In this case, Eq. (7) can still be used to count the number of resonances.

Using Eq. (6), for normal incident light, we calculate the upper limit of absorption enhancement factor as a function of $L/\lambda$ (Fig. 2b) when the structure has a square lattice. The discontinuous changes in Fig. 2b correspond to the emergence of new channels. In particular, when $\lambda > L$, there is always only a single channel independent of frequency. On the other hand, the number of resonances is frequency-dependent. As a result, the maximum enhancement factor increases quadratically as a function of frequency. In order to maximize the absorption, one should choose the periodicity be slightly smaller than the wavelength range of interest, since in this case one can obtain significant bandwidth where the upper limit of enhancement factor can exceed $4n^2$ (red region in Fig. 2b). We note that the ability to go beyond the Yablonovitch limit is only significant when the period is comparable or smaller than the wavelength. The upper limit for the absorption enhancement factor, on the other hand, approaches $4n^2$ for large period $L \gg \lambda$.

*Light-trapping in thin films.* When the thickness $d$ of the film is sufficiently small as compared to the wavelength, one can reach the single-mode regime where the film supports a single waveguide mode band for each of the two polarizations. In such a case, Eq. (7) is no longer applicable. Instead, the number of resonances in the frequency range of $[\omega, \omega + \delta\omega]$ can be calculated as:

$$M = 2 \times \frac{2\pi n_{wg}^2 \omega}{c^2} \left(\frac{L}{2\pi}\right)^2 \delta\omega \qquad (10)$$

where the first factor of 2 arises from counting both polarizations. (Here, to facilitate the comparison to the standard Yablonovitch limit, for simplicity we have assumed that the two polarizations have the same group index $n_{wg}$.) Notice that in this case the number of modes no longer explicitly depends upon the thickness $d$ of the film.

In order to highlight the effect of such strong light confinement, we choose the periodicity to be a few wavelengths, in which case the number of channels can still be calculated using Eq. (8). As a result we obtain the upper limit for the absorption enhancement factor:

$$F = 2 \times 4n_{wg}^2 \frac{\lambda}{4n_{wg}d} V \qquad (11)$$

where the factor $V = \frac{\alpha_{wg}}{\alpha_0}$ characterizes the overlapping between the profile of the guided mode and the absorptive active layer. $\alpha_{wg}$ and $n_{wg}$ are the absorption coefficient and group index of the waveguide mode respectively.

We note that Eq. (11) in fact becomes $4n^2$ in a dielectric waveguide of $d \approx \lambda/2n$. Therefore, reaching a single mode regime alone is not sufficient to overcome the Yablonovitch limit. Instead, to achieve the full benefit of nanophotonics, one must either confine the guided mode to a deep-subwavelength scale, or enhance the group index to be substantially larger than the refractive index of the active material, over a substantial wavelength range. Below, in the numerical example, we will design a geometry that simultaneously satisfies both these requirements.

*Numerical demonstration.* Guided by the theory above, we now numerically demonstrate a nanophotonic scheme with an absorption enhancement factor significantly exceeding $4n^2$. We consider a thin absorbing film with a thickness of 5nm (Fig. 3a), consisting of a material with a refractive index $n_L = \sqrt{2.5}$ and a wavelength-independent absorption length of 25μm. The film is placed on a mirror that is approximated to be a perfect electric conductor. Our aim here is to highlight the essential physics of nanophotonic absorption enhancement. The choice of material parameters therefore represents a simplification of actual material response. Nevertheless, we note that both the index and the absorption strength here are characteristic of typical organic photovoltaic absorbers in the weakly absorptive regime [21], while enhancing absorption in such a nanoscale layer is important for organic absorbers given their short exciton diffusion lengths [22-23].

In order to enhance the absorption in the active layer, we place a transparent cladding layer ($n_H = \sqrt{12.5}$) on top of the active layer. Such a cladding layer serves two purposes. First, it enhances density of state. The overall structure supports a fundamental mode with group index $n_{wg}$ close to $n_H$, which is much higher than that of the absorbing material. Second, the index contrast between active and cladding layer provides nanoscale field confinement. Fig. 3b shows the fundamental waveguide mode. The field is highly concentrated in the low-index active layer, due to the well-known slot-waveguide effect [24]. Thus, the geometry here allows the creation of a broad-band high-index guided mode, with its energy highly concentrated in the active layer, satisfying the requirement in Eq. (11) for high absorption enhancement.

In order to couple incident light into such nanoscale guided modes, we introduce a scattering layer with a periodic pattern on top of the cladding layer, with a periodicity $L$ much larger than our wavelength ranges of interest. Each unit cell consists of a number of air grooves. These grooves are oriented along different directions to ensure that scattering strength does not strongly depend on the angles and polarizations of the incident light.

We simulate the absorption in the proposed structure by numerically solving Maxwell's equations (Fig. 4a. details of the simulation available in supplementary information). The device has a spectrally-averaged absorption enhancement factor of $F = 119$ (red line) for normally incident light. (All the absorption spectra and enhancement factors are obtained by averaging $s$ and $p$ polarized incident light.) This is well above the Yablonovitch limit for both the active material ($4n_L^2 = 10$) and the cladding material ($4n_H^2 = 50$). Moreover, the angular response is nearly isotropic (Fig. 4c,d). Thus such enhancement cannot be attributed to the narrowing of angular range in the emission cone, and instead is due entirely to the nanoscale field confinement effect.

Using our theory, we calculate the theoretical upper limit of light-trapping enhancement in this structure. The structure supports three waveguide modes, for each of which we calculate the overlapping factor and group index (details in supplementary information). For wavelength $\lambda = 500\text{nm}$, we obtain an upper limit of $F = 147$. The enhancement factor observed in the simulation is thus consistent with this predicted upper limit. The actual enhancement factor obtained for this structure falls below the calculated theoretical upper limit because some of the resonances are not in the strong over-coupling regime.

Finally, to illustrate the importance of nanoscale field confinement enabled by the slot-waveguide effect, we change the index of the material in the absorptive layer to a higher value of $n_H$. Such a structure does not exhibit the slot-waveguide effect. The average enhancement in this case is only 37, falling below the Yablonovitch limit of 50 (Fig. 4b).

*Conclusion.* We have developed a statistical coupled-mode theory for nanophotonic light trapping, and shown that properly designed nanophotonic structures can achieve enhancement factors that far exceed the conventional Yablonovitch limit. We have illustrated the theory using a dielectric thin film as an example. The basic result, i.e. Eq. (6), however, is generally applicable to any photonic structures, including in particular nanowire and plasmonic structures, provided that appropriate mode counting is carried out for these systems. Finally, the numerical results presented here show that there is substantial opportunity for nanophotonic light-trapping to greatly enhance PV performance using only low-loss dielectric components.

Acknowledgement: This work was supported by the Center for Advanced Molecular Photovoltaics (CAMP) (Award No KUSC1-015-21), made by King Abdullah University of Science and Technology (KAUST), and by DOE Grant No. DE-FG02-07ER46426.

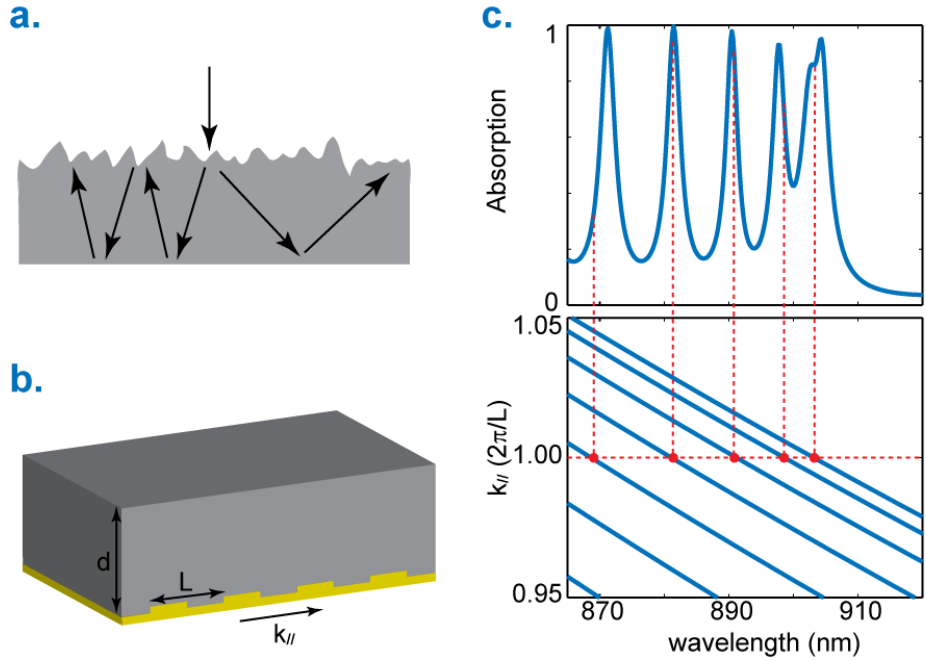

Fig. 1: a) Light trapping by randomly textured surface. b) Light-trapping with the use of a periodic grating on a back-reflector (yellow). $d = 2\mu m$. $L = 250nm$. The depth and width of the dielectric groove in the grating are 50nm and 175nm respectively. The dielectric material is crystalline silicon. c) Absorption spectrum (TM mode, normal incidence) and dispersion relation of waveguide modes for the structure in b). The dispersion relation is approximated by assuming $k_\perp = m\pi/d$, where $m = 1, 2, 3, \ldots$ is the band index. Resonances occur when $k_{//} = 2\pi/L$ (red dots).

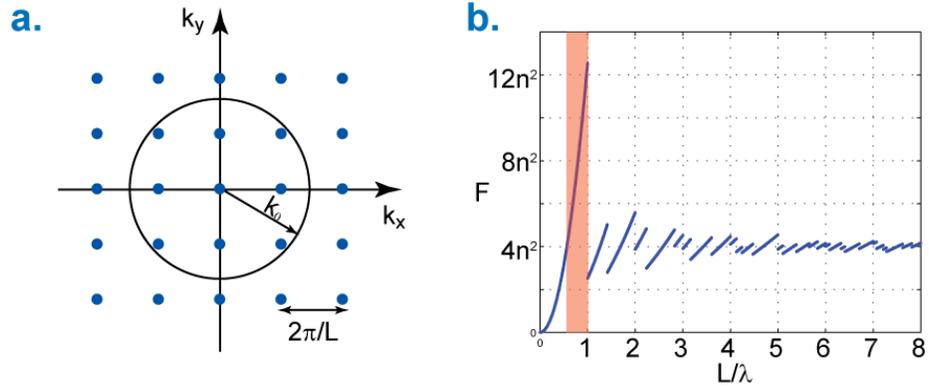

Fig. 2. a) Blue dots represent channels in the k-space. Channels in the circle correspond to free-space propagating modes. b) Theoretical upper limit of the absorption enhancement factor using a light trapping scheme where a square-lattice periodic grating structure is introduced into a thin film. Red area represents a spectral range where the upper limit of absorption enhancement factor $F$ is above $4n^2$.

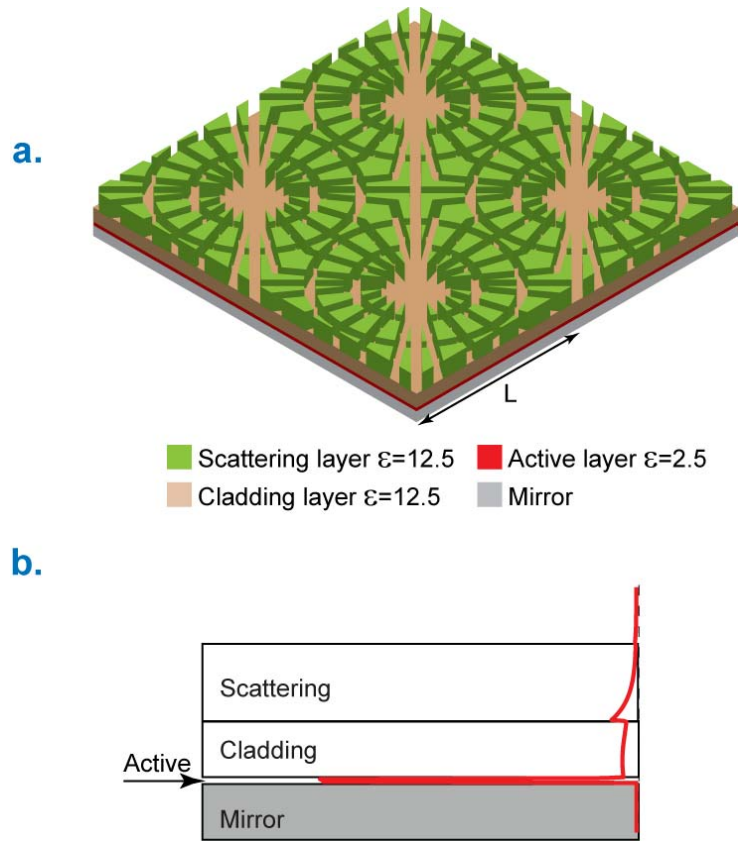

Fig.3 a) A nanophotonic light-trapping structure. The scattering layer consists of a square lattice of air groove patterns with periodicity $L = 1200$nm. The thicknesses of the scattering, cladding, and active layers are 80nm, 60nm, and 5nm respectively. The mirror layer is a perfect electric conductor. b) The profile of electric-field intensity for the fundamental waveguide mode. Fields are strongly confined in the active layer. To obtain the waveguide mode profile, the scattering layer is modeled by a uniform slab with an averaged dielectric constant.

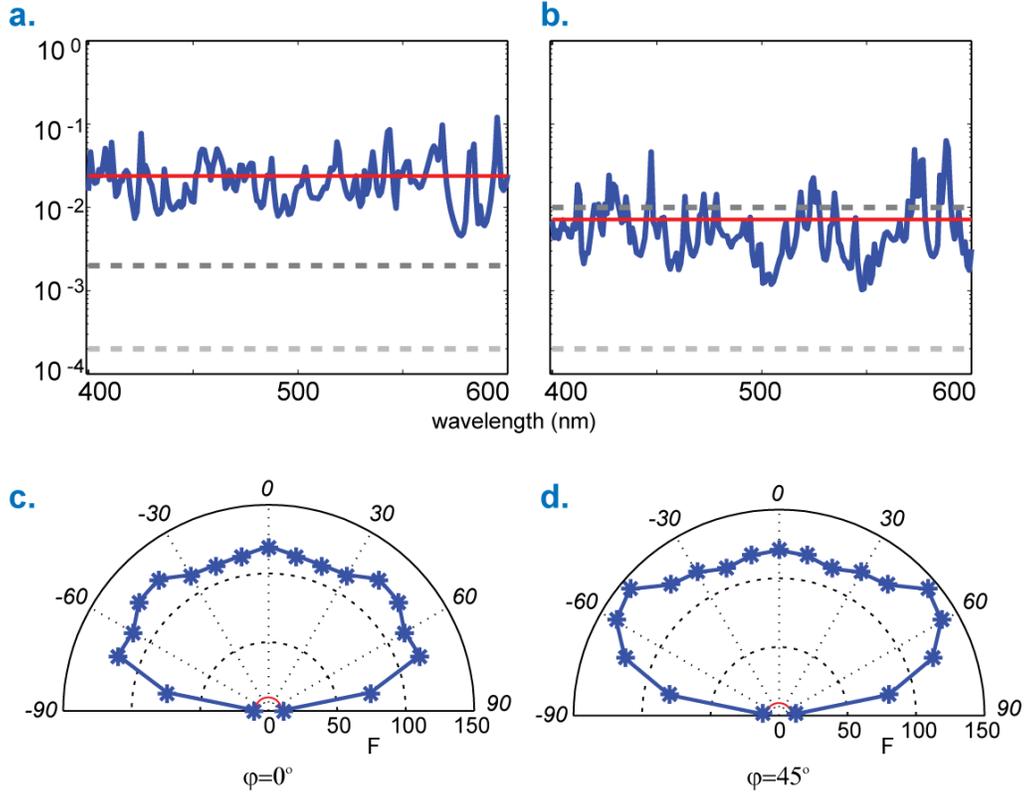

Fig. 4: a) Absorption spectrum for normally incident light for the structure shown in Fig. 3. The spectrally-averaged absorption (red solid line) is much higher than both the single pass absorption (light gray dashed line) and the absorption as predicted by the limit of $4n_L^2$ (dark gray dashed line). b) Absorption spectrum without light confinement. The structure is the same as that of a) except that the dielectric constant of the active layer is now the same as the cladding layer. The dark gray dashed line represents the absorption as predicted by the limit of $4n_H^2$. c, d) Angular dependence of the spectrally-averaged absorption enhancement factor for the structure in Fig. 3. Incident angles are labeled on top of the semi-circles. Incident planes are oriented at 0 (c) and 45 (d) degrees (azimuthal angles) with respect to the [10] direction of the lattice. The red circles represent the $4n_L^2$ limit.